# Research productivity: are higher academic ranks more productive than lower ones?[1]

*Giovanni Abramo*[a,b,*], *Ciriaco Andrea D'Angelo*[b] *and Flavia Di Costa*[b]

[a] National Research Council of Italy

[b] Laboratory for Studies of Research and Technology Transfer
School of Engineering, Department of Management
University of Rome "Tor Vergata"

**Abstract**

This work analyses the links between individual research performance and academic rank. A typical bibliometric methodology is used to study the performance of all Italian university researchers active in the hard sciences, for the period 2004-2008. The objective is to characterize the performance of the ranks of full, associate and assistant professors, along various dimensions, in order to verify the existence of performance differences among the ranks in general and for single disciplines.

---

# 1. Introduction

The science of research evaluation has developed rapidly over recent decades. Development has been particularly intense due to the general trend of many industrialized countries to initiate national research assessment exercises. Such exercises are intended to deal with the demands for greater accountability and improved allocative efficiency in funding of research institutions. Scholars and practitioners are developing measurement techniques and methodologies that are ever more robust and trustworthy, for various levels of application: macro (entire nations), meso (individual institutions) and micro (single researchers). There has been a flowering of studies and research concerning evaluation at the level of the individual, particularly at the methodological level, generally focused on single performance indicators. The motives for evaluation vary, from stimulating greater research productivity, to service in selective funding, to reduction of asymmetry between suppliers and users of new knowledge. Whatever its purpose, the evaluation must be as accurate as possible. Among the various factors that could justify differing performance among scientists in the same field of research (seniority, localization, capital available, etc.), the dimension that we explore in this study is the rank of the scientist in his or her research career path. In theory, the scientist's rank should reflect his/her demonstrated level of performance and future prospects. When the performance of a low-ranking scientist equals or exceeds the average of scientists at a higher level then advancement should be possible.

The literature seems to indicate that performance of the upper ranks of scientists is greater than that of others. Blackburn et al. (1978) surveyed a sample of faculty in US four-year colleges and universities in an attempt to identify factors affecting scholarly productivity in four disciplines: biological sciences, humanities, physical sciences, and social sciences. The study was conducted without taking account of the varying publication and citation intensity among scientific fields. The results observed that full professors publish at a higher average rate than associate professors and research staff.

More recently, Ventura and Mombrù (2006) conducted a study concerning publication and citation profiles of full and associate professors at the School of Chemistry of the Universidad de la República, Uruguay. The study is based on a very limited field of observation and, as the above one, ignores field-standardization. However the results seemed to confirm that full professors have greater averages than associate professors for number of papers and citations, and that this difference is significantly different.

One might explain this apparent phenomenon by arguing that higher ranks of academics generally hold greater seniority and thus possess greater experience and a better body of personal knowledge and competencies. However the correlation between seniority and productivity should not be so readily accepted as fact. It is true that experience increases with seniority, as we would also expect for an individual's social network and "status effect", and thus the ability to draw on tangible and intangible resources for research activity. However age may negatively influence ability to produce original research ideas and hypotheses. A popular belief states that science is a young person's game: Carl F. Gauss was 18 when he developed least-squares; Charles Darwin was 29 when he developed the concept of natural selection; Albert Einstein was 26 when he formulated the theory of relativity and Isaac Newton was 24 when he began work on universal gravitation (Cole, 1979).

Empirical studies on this theme do not always agree in their conclusions. Both Costas et al. (2010) and Shin and Cummings (2010) report that age has negative effects on research performance. The two studies in question deal respectively with scientists at the Spanish National Research Council and at institutions in South Korea. A study by Lissoni et al. (2011) concerning roughly 3,600 French and Italian physicists showed that the age of academics is negatively correlated to both number of publications and impact. Kyvik and Olsen (2008) do not achieve such clear results, and nor do Carayol and Matt (2006), who conclude "the effects of age on productivity remain an open question". Gingras et al. (2008) examine a sample of university professors in Quebec who had published at least one article between 2000 and 2007, and show that while average scientific impact per article decreases linearly until about age 50, the average number of articles in highly cited journals and among highly cited articles rises continuously until retirement.

Less studied, but as interesting as age, are the potential links between academic rank and research productivity. Lower academic ranks normally correspond to lower salaries, and thus to expectations of less output compared to higher ranks. In comparing scientists, it would thus be more appropriate to distinguish them and compile classifications for each of the ranks, rather than giving simple overall rankings. In comparisons of research institutions and their internal staff units it should also be advisable to take account of their different rank compositions. The current work proposes to back up these policy recommendations through an empirical analysis of all research personnel in the hard sciences that hold formal roles in Italian universities. The research performance will be evaluated for each scientist, characterized by academic rank, through a purely bibliometric approach using indicators for production (number of publications) and impact (citations). The analysis will be conducted at the level of the entire population (30,677 scientists), rather than a sample, avoiding the limits of inferential analysis and producing robust findings; and the performance evaluation methods allow for field standardization, thus avoiding distortions due to varying citation fertility among different research fields (Abramo et al., 2008). The current study is not intended to investigate the possible causes of the differences encountered, rather to provide more insights to the management of research assessment exercises.

The next section of the work describes the field of observation. Section 3 presents the data used and the bibliometric indicators applied. Section 4 presents the results of the analyses for the various disciplines. In the final section the authors provide their conclusions and indicate further possible investigations suggested by the findings.

## 2. Field of observation

In Italy, the Ministry of Education, Universities and Research (MIUR) recognizes a total of 95 universities, with the authority to issue legally-recognized degrees. With only rare exceptions these are public universities. These universities are largely financed through non-competitive allocation from the MIUR, although this share of income is decreasing. as seen in the reduction from 61.5% in 2001 to 55.5% in 2007 (MIUR, 2010). Up to 2009, the core funding by government was input oriented, meaning that it was distributed to universities in a manner intended to satisfy the needs for resources of each and all, in function of their size and activities. It was only following the first national research evaluation exercise (VTR), conducted between 2004 and 2006, that a minimal share, equivalent to 7% of MIUR financing[2], was attributed in function of the research evaluation and of teaching quality. All new personnel enter the university system through public examinations, and career advancement also requires such public examinations. Salaries are regulated at the nationally centralized level and are calculated according to role (administrative, technical, or professorial), rank within role (for example: assistant, associate or full professor), and seniority. No part of the salary for professors is related to merit: wages increase annually according to parameters set by government. All professors are contractually obligated to carry out research, thus there is no development of research and teaching universities. Differences between full, associate and assistant professors occur in salary and informal hierarchical dependency. Form an organizational standpoint higher ranks do not have any power over lower ranks. In practice, however, higher ranks have an enormous power over lower ranks because career advancement is often determined by full professors, who lobby together to have their own candidates win the public examinations[3].

The period of observation of the study is from 2004 to 2008. The entire Italian university population over this period consisted of approximately 66,000 scientists. Under the Italian university system, each researcher is classified as belonging to only one scientific field, called a "scientific disciplinary sector" (SDS), of which there are 370 in all. SDSs are grouped in 14 disciplines called "university disciplinary areas" (UDAs)[4]. This study examines the nine UDAs that deal with the hard sciences[5] and, within these, only those SDSs in which 50% of the researchers achieved at least one publication during the period observed (186 of a total 205 SDSs). For greater significance, observation was also limited to those scientists who were faculty members over the entire five-year period. Thus the dataset for the analysis included 30,677 scientists, sorted in the UDAs and academic ranks as indicated in Table 1.

The table shows that the division of the scientists by rank is almost uniform, with 35.1% full professors, 34.1% associate and 30.9% assistants. The full professors are more numerous than other ranks in five out of nine UDAs. Assistant professors are more numerous only in the Medicine UDA. The greatest difference between the ranks is found in Industrial and information engineering, where full professors (44.7%) exceed assistant professors (20.7%) by a full 24 percentage points.

The average age of the scientists in relation to their academic rank is presented in Table 2. The highest overall average age for a UDA is observed for Medicine (54), while the minimum value (48) occurs in Industrial and information engineering. Concerning academic rank, the average ages are 60 for full professors, 53 for associates and 45 for assistants. The highest average age for full professors (62) is in Physics and the minimum (57) is in Mathematics and computer science. For associate professors, the maximum average age (56) is in Medicine and the minimum (49) is in Industrial and information engineering. Finally, for assistant professors the highest average age (49) is in Medicine and the minimum (39) is in Industrial and information engineering.

---

[2] Since MIUR financing composes 55.5% of the total, the share that is distributed on the basis of the VTR represents 3.9% of total income.
[3] Evaluation committees in public examinations are made of higher ranks professors.
[4] http://attiministeriali.miur.it/UserFiles/115.htm last accessed on April 28, 2011
[5] Mathematics and computer sciences; Physics; Chemistry; Earth sciences; Biology; Medicine; Agricultural and veterinary sciences; Civil engineering; Industrial and information engineering.

| UDA | SDS | Professors | | | Total |
|---|---|---|---|---|---|
| | | Full | Associate | Assistant | |
| Mathematics and computer science | 9 | 1,056 (37.2%) | 1,035 (36.5%) | 744 (26.2%) | 2,835 |
| Physics | 8 | 847 (37.1%) | 890 (39.0%) | 544 (23.8%) | 2,281 |
| Chemistry | 12 | 1,013 (35.8%) | 1,067 (37.7%) | 752 (26.6%) | 2,832 |
| Earth sciences | 12 | 385 (35.0%) | 427 (38.8%) | 288 (26.2%) | 1,100 |
| Biology | 19 | 1,562 (34.9%) | 1,491 (33.3%) | 1,427 (31.9%) | 4,480 |
| Medicine | 49 | 2,647 (27.9%) | 2,925 (30.8%) | 3,910 (41.2%) | 9,482 |
| Agricultural and veterinary sciences | 28 | 941 (39.4%) | 775 (32.5%) | 671 (28.1%) | 2,387 |
| Civil engineering | 7 | 455 (40.4%) | 403 (35.8%) | 269 (23.9%) | 1,127 |
| Industrial and information engineering | 42 | 1,858 (44.7%) | 1,435 (34.6%) | 860 (20.7%) | 4,153 |
| Total | 186 | 10,764 (35.1%) | 10,448 (34.1%) | 9,465 (30.9%) | 30,677 |

*Table 1: Research personnel in Italian university hard sciences sorted by University Disciplinary Area and academic rank; data 2004-2008*

| UDA | Professors | | | Average |
|---|---|---|---|---|
| | Full | Associate | Assistant | |
| Mathematics and computer science | 57 | 52 | 42 | 49 |
| Physics | 62 | 54 | 45 | 53 |
| Chemistry | 61 | 53 | 42 | 51 |
| Earth sciences | 61 | 54 | 45 | 53 |
| Biology | 60 | 52 | 45 | 51 |
| Medicine | 61 | 56 | 49 | 54 |
| Agricultural and veterinary sciences | 59 | 51 | 43 | 50 |
| Civil engineering | 61 | 54 | 46 | 52 |
| Industrial and information engineering | 58 | 49 | 39 | 48 |
| Total | 60 | 53 | 45 | 52 |

*Table 2: Average age of research personnel in Italian university hard sciences, by UDA and academic rank; data 2004-2008*

### 3. Bibliometric indicators and data

Research performance of each individual scientist will be measured using three bibliometric indicators, which reflect the most common questions concerning potential differences between scientists of different ranks: i) does a full professor publish more or less than associate and assistant professors; ii) is the average impact of each publication greater or lesser; iii) is the overall impact greater or lesser?

The indicators are thus defined:
- $N_p$: number of publications authored by a scientist,
- Quality Index (QI): average of standardized citations[6] for the publications authored by a scientist, where citations of each publication are standardized by dividing by the median[7] of citations[8] for all Italian publications of the same year and same ISI subject category.
- Fractional Scientific Strength (FSS): sum of the standardized citations for

---
[6] Including self-citations.
[7] Standardization of citations with respect to median value rather than to the average (as frequently observed in literature) is justified by the fact that distribution of citations is highly skewed in almost all disciplines.
[8] Observed as of 30/06/2009.

publications authored by a scientist, each divided by the number of co-authors of the publication[9].

For the hard sciences, the literature gives ample justification for the choice of the bibliometric approach, reasoning that: i) scientific publications are a good proxy of overall research output (Moed et al., 2008); and ii) citations are a good proxy of impact on scientific advancement, notwithstanding the possible distortions inherent in this indicator (Glänzel, 2008).

The data used are taken from the ORP (Observatory of Italian Public Research[10]) a database that the authors derive from the Thomson Reuters Web of Science (WoS). Beginning from the raw data indexed in the WoS, then applying a complex algorithm for reconciliation of the authors' affiliation and disambiguation of the true identity of the authors, each publication (articles, reviews, and conference proceedings) is attributed to the Italian university scientists that produced it, with an error of less than 5% (D'Angelo et al., 2010).

## 4. Results

This section presents the results for the three analyses of interest: rate of activity per academic rank, scientific performance and degree of concentration of performance related to academic rank.

### 4.1 Rates of activity for academic ranks

Of the researchers in the dataset, 86.2% are authors of at least one publication listed in the WoS over the period under observation. Table 3 shows the number of active researchers in each UDA and academic rank as absolute value and percentage of total. The highest percentage of active scientists is in Chemistry, both for the overall average and in each of the three ranks, with the rates ranging from 98.8% of full professors to 94.5% of assistants. Conversely, the lowest percentages of active personnel for all three ranks are all found in Civil engineering, where the top performing rank is full professor (83.7%) and the lowest performance is for assistant professors (64.7%). Globally, 91.6% of full professors, 86.0% of associates and 80.2% of assistant professors are active. In addition, in all UDAs, the highest percentage of actives is for full professors and the lowest percentage is for assistant professors, the sole exception being in Physics, where the associate professors are less active than assistants.

---

[9] For life sciences, different weights are given to each co-author according to his/her position in the list and the character of the co-authorship (intra-mural or extra-mural). If first and last authors belong to the same university, 40% of citations are attributed to each of them; the remaining 20% are divided among all other authors. If the first two and last two authors belong to different universities, 30% of citations are attributed to first and last authors; 15% of citations are attributed to second and last author but one; the remaining 10%.are divided among all others.

[10] www.orp.researchvalue.it last accessed on April 28, 2011

| UDA | Professors | | | Total |
|---|---|---|---|---|
| | Full | Associate | Assistant | |
| Mathematics and computer science | 906 (85.8%) | 799 (77.2%) | 548 (73.7%) | 2,253 (79.5%) |
| Physics | 796 (94.0%) | 789 (88.7%) | 490 (90.1%) | 2,075 (91.0%) |
| Chemistry | 1,002 (98.9%) | 1,017 (95.3%) | 711 (94.5%) | 2,730 (96.4%) |
| Earth sciences | 354 (91.9%) | 355 (83.1%) | 226 (78.5%) | 935 (85.0%) |
| Biology | 1,482 (94.9%) | 1,351 (90.6%) | 1,249 (87.5%) | 4,082 (91.1%) |
| Medicine | 2,475 (93.5%) | 2,472 (84.5%) | 2,901 (74.2%) | 7,848 (82.8%) |
| Agricultural and veterinary sciences | 845 (89.8%) | 674 (87.0%) | 570 (84.9%) | 2,089 (87.5%) |
| Civil engineering | 381 (83.7%) | 289 (71.7%) | 174 (64.7%) | 844 (74.9%) |
| Industrial and information engineering | 1,620 (87.2%) | 1,241 (86.5%) | 723 (84.1%) | 3,584 (86.3%) |
| Total | 9,861 (91.6%) | 8,987 (86.0%) | 7,592 (80.2%) | 26,440 (86.2%) |

*Table 3: Italian university scientists with at least one publication over the period 2004-2008, by UDA and academic rank*

The researchers that obtained at least one citation (non-nil FSS) are presented in Table 4, as absolute value and percentage of total per UDA. Over the period examined, only 79.4% of scientists received at least one citation, compared to the 86.2% that have at least one publication. This type of difference (percentage with citations compared to percentage with "active" publication) is somewhat less for the specific case of full professors, although there are variations among the full professors for the individual UDAs. The extremes are for Chemistry, which shows the minimal difference (0.8 percentage points) and Civil engineering, with the maximum (18.9 percentage points). Considering all the ranks, the greatest differences are almost always for the assistant professors, with the exceptions being the two cases of Physics and Chemistry, where associate professors have the greatest difference. Thus we can conclude that, for full professors, there is less of a gap between percentage of personnel who have at least one publication and percentage that have at least one citation.

Focusing only on those scientists that received at least one citation, we see that the percentages fall in a range from 64.8% for Civil engineering to 98.1% for Chemistry, for full professors. Lesser percentages are seen in the other academic ranks: 55.1%-93.8% for associate professors and 50.2%-93.5% for assistants, with the extremes of the range again being for Civil engineering and Chemistry.

| UDA | Professors | | | Total |
|---|---|---|---|---|
| | Full | Associate | Assistant | |
| Mathematics and computer science | 803 (76.0%) | 652 (63.0%) | 442 (59.4%) | 1,897 (66.9%) |
| Physics | 770 (90.9%) | 747 (83.9%) | 468 (86.0%) | 1,985 (87.0%) |
| Chemistry | 994 (98.1%) | 1,001 (93.8%) | 703 (93.5%) | 2,698 (95.3%) |
| Earth sciences | 326 (84.7%) | 321 (75.2%) | 201 (69.8%) | 848 (77.1%) |
| Biology | 1,461 (93.5%) | 1,301 (87.3%) | 1,188 (83.3%) | 3,950 (88.2%) |
| Medicine | 2,402 (90.7%) | 2,365 (80.9%) | 2,677 (68.5%) | 7,444 (78.5%) |
| Agricultural and veterinary sciences | 741 (78.7%) | 585 (69.0%) | 476 (70.9%) | 1,802 (75.5%) |
| Civil engineering | 295 (64.8%) | 222 (55.1%) | 135 (50.2%) | 652 (57.9%) |
| Industrial and information engineering | 1,397 (75.2%) | 1,075 (74.9%) | 612 (71.2%) | 3,084 (74.3%) |
| Total | 9,189 (85.7%) | 8,269 (79.1%) | 6,902 (72.9%) | 24,360 (79.4%) |

*Table 4: Italian university scientists with at least one citation over the period 2004-2008, per UDA and academic rank*

## 4.2 Differences in performance for academic ranks

To study the links between academic rank and scientific performance, especially to verify which academic rank registers the best performance, the study provided for four micro-analyses: i) analysis of average percentile for the indicators $N_p$, FSS and QI, by UDA and academic rank (section 4.2.1); ii) analysis of the average position of full professors and assistant professors using the casual variables sequence criterion (section 4.2.2); iii) analysis of the level of concentration of performance (Section 4.2.3); iv) analysis of the incidence of the three ranks among top scientists of the Italian university system (section 4.2.4). The analyses were conducted by beginning from the level of the individual scientists under observation. A national ranking is provided for each scientist compared to other colleagues in the same SDS, for the each of the indicators used. The ranking, expressed as a percentile[11], permits a comparative analysis of scientists belonging to different fields and disciplines that are not uniform in terms of dimension or scientific prolificacy (Abramo and D'Angelo, 2011). By aggregating the data it is then possible to extend the comparison to the level of UDAs. For this, calculation is made of the average percentile rank for all scientists in each academic rank, for each UDA.

### 4.2.1 Differences in percentiles for performance indicators

Research performance was measured along three dimensions: number of publications, average impact per publication, overall impact. Table 5 presents the differences in scientific production. At the aggregate level, full professors publish more (53.43) than associate professors (52.14), who publish more than assistants (46.70). At the level of the UDA, the assistant professors always average fewer publications than the other academic ranks and the full professors always publish more, with the exception of those in Industrial and information engineering, Chemistry and Biology. The maximum difference between the ranks is in Civil engineering (13.38) and the minimum difference is in Physics (3.71).

| UDA | Professors | | |
|---|---|---|---|
| | Full | Associate | Assistant |
| Mathematics and computer science | 52.45 | 49.78 | 44.90 |
| Physics | 52.79 | 52.66 | 49.08 |
| Chemistry | 53.33 | 55.11 | 48.70 |
| Earth sciences | 55.84 | 54.12 | 45.82 |
| Biology | 53.19 | 53.36 | 48.73 |
| Medicine | 54.34 | 51.05 | 45.98 |
| Agricultural and veterinary sciences | 54.74 | 52.74 | 48.16 |
| Civil engineering | 54.15 | 50.09 | 40.77 |
| Industrial and information engineering | 51.89 | 51.97 | 45.94 |
| Total | 53.43 | 52.14 | 46.70 |

*Table 5: Average percentile for indicator $N_p$ by UDA and academic rank*

---

[11] Comparison among all scientists of the same SDS permits calculation of national percentile of performance (0 being worst, 100 being best) for each indicator.

Table 6 presents results from the analysis for the indicator of total impact, FSS. The results of the preceding analysis are generally confirmed but the differences in performance are reduced. At the level of the individual UDAs, the average impact of assistant professors is always less than other ranks. Full professors always perform better except in Chemistry, but the difference from Associate professors is minimal for Biology and Industrial and information engineering. The greatest percentile difference between the faculty ranks is in Civil engineering (8.62) and the smallest difference is in Physics (1.88).

| UDA | Professors | | |
| --- | --- | --- | --- |
| | Full | Associate | Assistant |
| Mathematics and computer science | 48.20 | 42.92 | 41.24 |
| Physics | 51.33 | 50.34 | 49.45 |
| Chemistry | 52.10 | 53.43 | 50.21 |
| Earth sciences | 51.35 | 49.02 | 44.04 |
| Biology | 51.18 | 50.78 | 47.96 |
| Medicine | 52.19 | 48.88 | 43.85 |
| Agricultural and veterinary sciences | 48.86 | 46.80 | 45.66 |
| Civil engineering | 45.22 | 40.23 | 36.60 |
| Industrial and information engineering | 47.02 | 47.34 | 46.69 |
| Total | 50.07 | 48.46 | 45.28 |

*Table 6: Average percentile for indicator FSS by UDA and academic rank*

Table 7 presents the results of analysis for average impact of publications. The results confirm the preceding two analyses but with still lesser differences. Full professors have a higher average impact than all others in six of nine UDAs, while associates have highest impact in Chemistry and Biology and assistants excel in Industrial and information engineering. The maximum percentile difference begtween ranks is in Civil engineering (7.72 percentage points) and the minimum is in Physics (0.75).

| UDA | Professors | | |
| --- | --- | --- | --- |
| | Full | Associate | Assistant |
| Mathematics and computer science | 48.05 | 42.76 | 42.40 |
| Physics | 50.75 | 50.00 | 50.56 |
| Chemistry | 51.10 | 52.57 | 52.38 |
| Earth sciences | 50.67 | 49.04 | 45.51 |
| Biology | 50.37 | 50.41 | 48.89 |
| Medicine | 50.97 | 49.33 | 44.79 |
| Agricultural and veterinary sciences | 48.67 | 46.56 | 46.51 |
| Civil engineering | 45.24 | 40.23 | 37.52 |
| Industrial and information engineering | 46.59 | 47.22 | 47.93 |
| Total | 49.38 | 48.36 | 46.38 |

*Table 7: Average percentile for indicator QI by UDA and academic rank*

**4.2.2 Causal variables sequence criterion**

Differences in performance between academic ranks can also be calculated by applying the causal variables sequence criterion. Such analysis informs on which SDSs within a UDA an academic rank outperforms another one, on average. We present the example of the comparison between full and assistant professors, for the three performance indicators. Beginning with the performance ranking of each full professor (FP) versus assistant professor (AP) within each SDS, the distance between the ideal and effective case was measured. In analytical terms:

$$R_{FP-j}^{diff} = R_{FP-j}^{max} - R_{FP-j}^{eff}$$

$R_{FP-j}^{max}$ = sum of the ranks of full professors in sector j under the hypothesis of maximum differentiation*

$R_{FP-j}^{eff}$ = sum of the ranks of full professors in sector j

*"maximum differentiation" is understood as the situation in which the highest performing assistant professor is still ranked below the lowest full professor.*

The value $R_{FP-j}^{diff}$ therefore represents the "distance" for the ideal situation of maximum performance difference between academic ranks in favor of full professor. The same calculation is completed for assistant professor, and through comparison between $R_{FP-j}^{diff}$ and $R_{AP-j}^{diff}$ it can be determined which of the two populations, full professor or assistant professor, obtains a higher overall ranking. The simple sum of the data by SDS provides the overall view at the level of UDA. The results show that, for all three indicators, full professors show higher overall performance than that of assistant professors, in all UDAs considered.

Table 8 presents the numbers of SDS in which assistant professors register a higher overall ranking than full professors, on the basis of the causal variables sequence criterion. Assistant professors achieve higher average publication in only 8 out of the 176 SDSs examined. They have higher overall impact (FSS) in 6 SDSs and better average impact per publication (QI) in 11. The Industrial and information engineering UDA has the greatest concentration of those few SDSs where assistant professors have a better ranking for all three indicators.

| UDA | $N_p$ | FSS | QI |
|---|---|---|---|
| Mathematics and computer science | 0 out of 9 | 0 out of 9 | 0 out of 9 |
| Physics | 0 out of 7 | 0 out of 7 | 1 out of 7 |
| Chemistry | 0 out of 11 | 0 out of 11 | 1 out of 11 |
| Earth sciences | 1 out of 12 | 0 out of 12 | 1 out of 12 |
| Biology | 0 out of 19 | 0 out of 19 | 1 out of 19 |
| Medicine | 0 out of 46 | 0 out of 46 | 0 out of 46 |
| Agricultural and veterinary sciences | 1 out of 26 | 1 out of 26 | 2 out of 26 |
| Civil engineering | 0 out of 7 | 0 out of 7 | 0 out of 7 |
| Industrial and information engineering | 6 out of 39 | 5 out of 39 | 5 out of 39 |
| Total | 8 out of 176 | 6 out of 176 | 11 out of 176 |

*Table 8: Number of SDSs in which assistant professors outperform full professors, according to the causal variables sequence criterion*

**4.2.3 Concentration of performance**

The preceding sections present a comparative analysis of scientific performance in relation to academic rank, particularly in terms of average values. We also ask if academic ranks are characterized by different internal distributions of performance. For this, the levels of skewness of the performance distributions are of particular interest. Therefore, in this section we analyze the degree of concentration of performance, at the level of single SDS and at the UDA level, to understand if and when the scientific performance of university researchers is non-uniform and if there are significant differences between the UDAs and/or within the UDAs at the level of rank. The analysis is conducted using two methodologies: Gini coefficient[12]; ratio of bottom 40% to top 20%. For reasons of space, we present the results of the analyses only for the overall impact indicator, FSS (Table 9).

For Gini coefficient, values closer to 1 indicate greater concentration within the population considered and values closer to 0 indicate greater uniformity of performance among scientists. The average value of concentration at UDA level is obtained by weighting the individual indicators of concentration for each SDS by the share of scientists on staff in that SDS out of the total for the UDA. Thus, Gini coefficient for each UDA results as:

$$\frac{1}{Add_{UDA_{i,k}}} \sum_{j=1}^{n_{SDS}} Coeff\ Gini_{jk}\ Add_{jk}$$

$Add_{UDA_{i,k}}$ = total of scientists in UDA $i$ and rank $k$
$Add_{jk}$ = total of scientists in SDS $j$ and rank $k$
$Coeff\ Gini_{jk}$ = Gini coefficient for performance of scientists in SDS $j$ and rank $k$
$n_{SDS}$ = total number of SDSs for the $UDA_i$ considered

For FSS, the Gini coefficient falls between 0.5 and 0.8 and the differences between academic ranks within the UDAs are minimal. The Chemistry UDA has the least degree of concentration of performance, with values ranging between 0.542 (assistant professors) and 0.573 (associate professors). For full professors, values range from a maximum of 0.726 in Civil engineering to a minimum of 0.560 for Chemistry. For associate professors the range is between 0.750 (Civil engineering) and 0.573 (Chemistry). Finally, for assistant professors the values vary from 0.767 (Medicine) to 0.541 (Chemistry).

The Gini coefficient shows a tendency for concentration of performance in the various UDAs. For confirmation, the degree of concentration for performance in disciplines was also measured through the ratio of cumulative bottom 40% to top 20% performances. The calculation considers the two subgroups of personnel: "top scientists", meaning those placing in the top 20% of national rankings for FSS, and a second group of the bottom-ranking 40%. The aggregate measure of UDA is obtained by weighting the ratio of the cumulative bottom 40% to top 20% of performances, calculated for individual SDSs, by the share of scientists on staff in the SDS out of the total personnel in the UDA. The values obtained for this ratio are complementary to the Gini coefficient: those tending to 0 indicate greater concentration of performance and those tending to 1 indicate greater uniformity of performance.

---

[12] Here, Gini coefficient is a measure of inequality of research productivity: a value of 0 suggests that variation among scientists is nil; a value of 1 indicates maximal inequality.

|  | Professors | | | | | |
|---|---|---|---|---|---|---|
|  | Full | | Associate | | Assistant | |
| UDA | Gini | Bottom/Top | Gini | Bottom/Top | Gini | Bottom/Top |
| Mathematics and computer science | 0.655 | 0.040 | 0.724 | 0.015 | 0.707 | 0.013 |
| Physics | 0.615 | 0.082 | 0.658 | 0.048 | 0.625 | 0.060 |
| Chemistry | 0.560 | 0.144 | 0.573 | 0.114 | 0.542 | 0.131 |
| Earth sciences | 0.579 | 0.101 | 0.639 | 0.055 | 0.667 | 0.031 |
| Biology | 0.607 | 0.105 | 0.645 | 0.080 | 0.654 | 0.062 |
| Medicine | 0.635 | 0.081 | 0.689 | 0.037 | 0.767 | 0.011 |
| Agricultural and veterinary sciences | 0.642 | 0.063 | 0.645 | 0.060 | 0.679 | 0.025 |
| Civil engineering | 0.726 | 0.012 | 0.750 | 0.007 | 0.743 | 0.003 |
| Industrial and information engineering | 0.654 | 0.045 | 0.655 | 0.048 | 0.665 | 0.039 |

*Table 9: Analysis of performance concentration (FSS) by UDA and academic rank*

### 4.2.4 Analysis of top scientists

To complement the analyses of concentration, the performance distribution of the top scientists for each academic rank was also calculated, for each UDA. A "top scientist" of any academic rank is one who places in the top 20% of the national ranking for his/her SDS, for a particular indicator. This analysis permits quantification of the relative representation, among top scientists, of the different academic ranks. Table 10 shows the results for the indicator of overall impact, FSS.

|  | Professors | | |
|---|---|---|---|
| UDA | Full | Associate | Assistant |
| Mathematics and computer science | 39.1 (1.05) | 37.3 (1.02) | 23.6 (0.90) |
| Physics | 38.1 (1.03) | 39.1 (1.00) | 22.9 (0.96) |
| Chemistry | 36.0 (1.01) | 39.9 (1.06) | 24.0 (0.90) |
| Earth sciences | 37.2 (1.06) | 39.5 (1.02) | 23.3 (0.89) |
| Biology | 36.3 (1.04) | 34.7 (1.04) | 29.0 (0.91) |
| Medicine | 31.0 (1.11) | 30.9 (1.00) | 38.1 (0.92) |
| Agricultural and veterinary sciences | 40.8 (1.04) | 33.1 (1.02) | 26.1 (0.93) |
| Civil engineering | 41.5 (1.03) | 36.1 (1.01) | 22.4 (0.94) |
| Industrial and information engineering | 44.2 (0.99) | 34.7 (1.00) | 21.1 (1.02) |
| Total | 36.8 (1.05) | 34.7 (1.02) | 28.5 (0.92) |

*Table 10: Distribution of top scientists (%) by FSS per UDA among academic rank (concentration index in brackets)*

At the general level, distribution by academic rank is not particularly concentrated: 36.8% of top scientists hold the rank of full professor, compared to 34.7% for associate and 28.5% for assistant professors. However, the percentage differences of top scientists per academic rank and UDA should be interpreted with respect to their corresponding indices of concentration[13], seen in table 10. These clearly indicate that the greater incidence of full professors among top scientist is in part linked to their greater relative number. In fact, we consistently see values that are just over one for the full and associate professors, while the values for assistants are less than one, for all UDAs except Industrial and information engineering. Analysis of correlation between status for excellence (top scientists versus all others) and academic rank, by means of Pearson $\chi^2$ test, shows a statistically significant relationship only in Medicine (0.15%), and for the entire population (0.07%) without distinction for disciplinary area. From these results we can conclude that the personnel who excel in research are not necessarily those that hold higher academic rank.

---

[13] Concentration indexes represent a measure of association between two variables based on frequency data, varying around the neutral value of 1. For example, in Mathematics and computer science, the value of 1.05 derives from the ratio of two percentages: full professor-top scientists over all top scientists of the UDA (39.1%), divided by full professors over total academic staff of the UDA (37.2%).

## 5. Conclusions

While scholars of the discipline have continued to develop more robust and reliable methodologies for measurement of scientific performance, there has been less investigation concerning the potential correlation between productivity and academic rank. However the question is certainly relevant. Lower academic ranks normally correspond to lower salaries, and thus to expectations of less output compared to higher ranks. If higher academic rank corresponds to greater research performance then it would be more appropriate for evaluation techniques to distinguish individual scientists by rank and develop performance lists by such categories, in addition to overall rankings. In comparing research institutions and their internal organizational units it would also be necessary to take account of their diverse composition for rank of research staff. In this work we have confirmed the needs for such evaluation methodologies, through empirical analyses of all research personnel in the hard sciences, on faculty in Italian universities over the period 2004-2008. The research performance of each scientist was evaluated through a purely bibliometric approach, particularly through use of indicators for number of publications and relative impact (citations). From the analysis of the subdivision of scientists by academic rank in each UDA, we observe that the distribution is not pyramidal, but almost uniform, although slightly shifted in favor of full professors. The percentage of full professors with at least one publication over the five-year period is greater than that for the other two ranks. The same holds concerning percentages that have obtained at least one citation. For number of publications and relative impact, full professors show the best performance, followed by associates and assistants. Scientific performance is significantly concentrated for all three populations, but more so for associate and assistant professors than for full professors. Analysis restricted to "top scientists" shows that the three ranks are represented similarly in this subpopulation, although the figures for full professors are slightly greater. In reality, detailed investigation by discipline shows significant differentiations: for example, in Medicine the greater part of top scientists are found among the ranks of assistant professors. In general, even though full professors dominate at the numeric level, an attentive analysis at the level of index of concentration shows that in reality the greater incidence of full professors among top scientists is in part linked to their greater relative numerosity.

The current study was not intended to investigate the causes of the differences encountered, which could be the topic of subsequent research. Among the possible determinants, are worth considering age/experience, access to financial support, leadership of organizational units and research teams, network of collaborators, reputation among peers, and least but not last the informal hierarchical power typical of the Italian university system. In terms of policy indications, the results of the analyses lead the authors to recommend that, in research evaluation exercises, it would be appropriate to detail the analysis at the level of single scientist in function of their academic rank, and to also take into account the varying makeup of the research staff in the assessment units when evaluation is conducted at the level of research institutions and internal organizational units.